\documentclass[twocolumn,letter,10pt]{article}

\title{\vspace*{-4em}{\footnotesize \textbf{Proceedings of the IASTED
International Conference}\hfill \hspace*{1ex}}\\[-0.8em]
{\footnotesize \textbf{Parallel and Distributed Computing and
Systems}\hfill \hspace*{1ex}}\\[-0.8em]
{\footnotesize \textbf{November 3--6, 1999 MIT, Boston, USA}\hfill
\hspace*{1ex}}\\[3em]
\textbf{\Large Stream parallel skeleton optimization}\vspace{-1ex}}
\author{\normalsize{M. Aldinucci \& M. Danelutto}\vspace*{-1em}}
\date{\vspace*{-1ex}\normalsize{Department of Computer Science --
University of Pisa -- Italy \\  
Email: \texttt{\{aldinuc,marcod\}@di.unipi.it}}}

\usepackage{fullpage}
\usepackage{graphicx}
\usepackage{layout}

\newcommand{\skie}{\mbox{\sf SkIEcl}}
\newcommand{\skieenv}{\mbox{\sf SkIE}}
\newcommand{\pppl}{\mbox{\sf P3L}}
\newcommand{\farm}{\mbox{$\diamond$}}
\newcommand{\Farm}[1]{\mbox{$\diamond$}({#1})}
\newcommand{\pipe}{\mbox{$\mid$}}
\newcommand{\Pipe}[3]{\mbox{$#1_{#2}\pipe \ldots \pipe  #1_{#3}$}}
\newcommand{\seqcomp}{\mbox{$;$}}
\newcommand{\Seqcomp}[3]{\mbox{$#1_{#2}\seqcomp \ldots \seqcomp #1_{#3}$}}
\newcommand{\seq}{\texttt{seq}}

\newcommand{\rufreccia}[1]{\mbox{$\stackrel{\tiny #1}{\to}$}}
\newcommand{\llista}[3]{\mbox{$#1_{#2}, \ldots , #1_{#3}$}}
\newcommand{\llistapv}[3]{\mbox{$#1_{#2}; \ldots ; #1_{#3}$}}
\newcommand{\stage}{\mbox{$\iota$}}
\newcommand{\skel}{\mbox{$\sigma$}}
\newcommand{\semf}{\mbox{${\cal F}$}}
\newcommand{\normale}[1]{\mbox{$\overline{#1}$}}
\newcommand{\service}[1]{\mbox{$T_{s}(#1)$}}
\newcommand{\seqservice}[1]{\mbox{$\tin{#1} + \tout{#1} + \tseq{#1}$}}
\newcommand{\farmservice}[1]{\mbox{$\min\{\max\{\tin{#1}, 
\tout{#1}\}, \service{#1}\}$}}

\newcommand{\pipeservice}[3]{\mbox{$\max\{\service{#1_{#2}},\ldots,\service{#1_{#3}}\}$}}

\newcommand{\tin}[1]{\mbox{$T_{i}(#1)$}}
\newcommand{\tout}[1]{\mbox{$T_{o}(#1)$}}
\newcommand{\tseq}[1]{\mbox{$T_{seq}(#1)$}}
\newcommand{\frangia}[1]{\mbox{${\sf fringe}(#1)$}}
\newcommand{\stream}[1]{\mbox{$\langle$ #1 $\rangle$}}
\newcommand{\streaml}[3]{\mbox{$\langle {#1}_{#3},\ldots ,{#1}_{#2} \rangle$}}
\newtheorem{teo}{Statement}

\newcommand{\lista}[2]{\mbox{$#1_1,\ldots,#1_{#2}$}}

\renewcommand{\min}{\mbox{$\downarrow$}}
\renewcommand{\max}{\mbox{$\uparrow$}}

\begin{document}

\newpage
\maketitle

\paragraph{Abstract}
  We discuss the properties of the composition of stream parallel
  skeletons such as pipelines and farms.  By looking at the ideal
  performance figures assumed to hold for these skeletons, we show
  that any stream parallel skeleton composition can always be
  rewritten into an equivalent ``normal form'' skeleton composition, 
  delivering a service time which is equal or even better to the
  service time of the original skeleton composition, and achieving a
  better utilization of the processors used. The normal form is
  defined as a single farm built around a sequential worker code.
  Experimental results are discussed that validate this normal form.

\noindent
\textbf{Keywords:} skeletons, rewriting, stream parallelism.


\section{Introduction\vspace*{-1ex}}
\label{sec:intro}
\footnotetext{This work has been partially supported by the MOSAICO
  and PQE2000 italian projects.} Skeleton based programming models
  represent an interesting subject in the field of parallel
  programming models.  Cole introduced the skeleton concept in the
  late 80's \cite{cole-th}.  Cole's skeletons represented parallelism
  exploitation patterns that can be used (instantiated) to model
  common parallel applications.  Later, the skeleton concept
  evolved. Different authors recognized that skeletons can be used as
  constructs of an explicitly parallel programming language, actually
  as the exclusive way in these languages to express parallel
  computations \cite{darlington-parle,fgcs-firenze,bacs}.  Recently,
  the skeleton concept evolved again, and became the coordination
  layer of structured parallel programming environments
  \cite{Darlington1996,skie-parcom,skie-marcov}.  In all cases, however, a
  skeleton can be understood as a higher order function taking one or
  more other skeletons or portions of sequential code as parameters,
  and modeling a parallel computation out of them.  Different
  parameters passed to the same skeleton give different parallel
  applications, all exploiting the same kind of parallelism: the one
  encapsulated by the skeleton used.

In most cases, in order to implement skeletons efficiently on 
parallel architectures, compiling tools based on the concept of
the implementation template have been developed
\cite{cole-th,orlando-grosso,libro-susanna}.  The implementation
templates are known, efficient and parametric ways (process networks,
actually) of implementing a skeleton on a given target architecture.
Therefore, the process of generating parallel code from a skeleton
source code can be performed by looking at the skeleton nesting in
the source code and deriving a corresponding nesting of (suitably
instantiated) implementation templates
\cite{anacleto-australia,libro-susanna}.
Furthermore, due to the fact that the skeletons have a clear
functional and parallel semantics, different rewriting techniques have
been developed (or derived from other functional programming contexts
\cite{bird-grosso,gorlatch97}) that allow skeleton programs to be
transformed/rewritten into equivalent ones, achieving different
performance when implemented on the target architecture
\cite{darlington-parle}. These rewritings can also be driven by some
kind of analytical performance models associated with the
implementation templates of the skeletons, in such a way that only
those rewritings leading to more efficient implementation of the
skeleton code at hand are considered \cite{modelli-aldinuc-coppola}.

The skeletons considered by the structured parallel programming
community range from a small number of very simple, general purpose
skeletons \cite{cole-th,darlington-parle,orlando-grosso} to a large
number of very specific, application oriented skeletons
\cite{kessler-svizzera}. Usually, the skeleton set comprises both
data parallel and control parallel skeletons. The data parallel
skeletons model parallel computations whose parallel activities come
from the computation of a single data item, whereas control parallel
skeletons model parallel computations whose parallel activities come
from the computation of different, independent data items. In turn,
stream parallel skeletons are those control parallel skeletons
exploiting parallelism in the computation of streams of results out of
streams of (independent) data items.

In this paper, we discuss the properties of programs derived by
nesting stream parallel skeletons and implemented via implementation
templates. Both the skeletons and the implementation templates
considered are those used in \pppl\ and \skie\ 
\cite{orlando-grosso,skie-userman-98}. 
In particular, we will show how arbitrary compositions of stream
parallel skeletons, including pipelines and farms, can always be
rewritten as farms of sequential code and we will prove that this
second form delivers a better service time.


\section{The skeleton framework\vspace*{-1ex}}
\label{sec:skeletons}
We consider a skeleton framework containing only
\textit{stream parallel} skeletons, i.e. we do not take into account
data parallel skeletons. In particular, we consider skeletons modeling
pipeline and farm parallelism as well as sequential composition of
sequential computations, similar to the ones used in
\cite{darlington-parle,orlando-grosso}. 
%

Stream parallel skeletons exploit parallelism in the computation of
streams of results from streams of input data.
By data stream we mean an ordered, finite collection of homogeneous
(i.e., having the same type) data items, possibly being available at
different times. We denote a data stream with data items
\llista{x}{1}{n} by \stream{$x_n,\ldots,x_1$} and we assume that data
item $x_i$ is available immediately after data item $x_{i-1}$ was
available and immediately before $x_{x+1}$ is be available.
When computing any stream parallel skeleton \skel\ onto a data stream
\streaml{x}{1}{n}, the computation of any result data item appearing
onto the output stream \streaml{y}{1}{n} is independent of the
computation of any other result data item. 
In other words, according to the usual definition of pipeline and farm
skeletons (such as the one given by
\cite{darlington-parle,orlando-grosso}), our stream parallel skeletons
denote \textit{stateless} computations.
Therefore, given any skeleton program \skel\ an input data stream
\stream{$x_n,\ldots,x_1$} and the corresponding output data stream
\stream{$y_n,\ldots,y_1$}, it holds that $\forall i \in [1,n]\ y_i =
\semf(x_i)$ provided that $\semf(x_i)$ is the function computed by
skeleton \skel.

In the following paragraphs, we give the functional and parallel
semantics of the skeletons included in our set. The functional
semantics (denoted in the following by \semf) just denotes the results
computed by each skeleton, and it will be used to show that different
skeleton programs (i.e. programs exploiting different skeleton
nestings) compute the same results. The functional semantics of our
skeletons will be given by formally defining the function \semf.
The (informal) parallel semantics, instead, denotes the parallelism
exploitation patterns of the skeletons and will be used to derive the
analytical cost models of Section \ref{sec:performance}, that, in
turn, will be used to prove the effectiveness of normal form, stream
parallel skeleton programs discussed in Section
\ref{sec:collapsing}. For each skeleton, the informal parallel
semantics will be given by discussing the parallelism exploitation
pattern used to implement the skeleton.
Of course, both functional and parallel semantics of a skeleton must
be taken into account in order to understand the skeleton
peculiarities.

%
%
%
%
%
%
%
%
%
%
%

\textbf{Sequential skeleton} The sequential skeleton (denoted by the
keyword \texttt{seq}) simply transforms a sequential portion of code,
written in some language $l$ in a skeleton that can be used as a
parameter of other skeletons, such as the pipeline, the farm and the
sequential composition ones.
We assume that a function ${\cal H}_l$ exists, for any sequential
language $l$, such that $ {\cal H}_l [ \texttt{prog} ] = \phi : \alpha
\to \beta $\footnote{$f:\alpha \to \beta$ denotes that
  function $f$ has type $\alpha \to \beta$.}, where $\phi$ is the
function computed by the sequential fragment of code \texttt{prog},
the type of input data processed by $\phi$ is $\alpha$ and the type of
output data produced by $\phi$ is $\beta$.
Therefore, the function computed by a sequential skeleton is (its
functional semantics is defined by): $
\semf[\texttt{seq}(\texttt{prog})] = {\cal H}_l [ \texttt{prog} ] =
\phi $
No parallelism is exploited while computing a sequential skeleton,
i.e. we assume that the whole computation of
$\semf[\texttt{seq}(\texttt{prog})]$ is performed sequentially on a
single processing element.

\textbf{Sequential composition} The sequential composition skeleton
(denoted by the infix operator ``\seqcomp'') models the sequential
composition of sequential skeletons. 
Therefore, the function computed by the skeleton:
$\Seqcomp{\stage}{1}{k}$ turns out to be\footnote{$\circ$ denotes the
usual function composition $(g\circ f)(x) = g(f(x))$.}: $ \semf[
\stage_1 \seqcomp \ldots \seqcomp \stage_k] = \semf[\stage_k] \circ
\ldots \circ \semf[\stage_1]$ provided that each $\stage_i =
\seq(\texttt{prog})$\footnote{In the following, we will denote by
$\stage_i$ sequential skeletons such as $\seq(\texttt{prog})$.}, and
the type of the resulting function is $ \alpha_1 \to \alpha_{k+1} $
(provided that $\forall i\ \semf(\stage_i): \alpha_{i} \to
\alpha_{i+1})$.
No parallelism is exploited when evaluating the sequential composition
skeleton, i.e. we assume that the whole computation of $ \semf[
\stage_1 \seqcomp \ldots \seqcomp \stage_k]$ is performed on a single
processing element.

\textbf{Pipeline} The pipeline skeleton (denoted by the infix operator
``\pipe'')  models function composition.  Whenever a function
$f$ of the input data has to be computed, such that the function can
be expressed as the composition of some functions $f_1, \ldots, f_k$,
i.e.  $ f = f_k \circ \ldots \circ f_1 $, a pipeline skeleton can be
used.
Given a pipeline skeleton, such as $\skel_1 \pipe \ldots \pipe
\skel_k$ the function computed by the skeleton is $\semf[\skel_1 \pipe
\ldots \pipe \skel_k] = \semf[\skel_k] \circ \ldots \circ
\semf[\skel_1]$ and the type of the resulting function is $ \alpha_1
\to \alpha_{k+1} $ (provided that $\forall i\ \semf(\skel_i):
\alpha_{i} \to \alpha_{i+1})$.
Parallelism is exploited in the computation of a pipeline skeleton as
the computations relative to the different stages on different data
items can be performed in parallel.
In principle, given a $k$ stage pipeline operating on
an input data stream \streaml{x}{1}{n}, the computation of stage $i$
(i.e. the computation of $\semf[\skel_i]$) relative to the input data
item $x_j$, can be performed in parallel with the computations of
stages $i' \in (1,k)$ s.t. $i'\neq i$ relative to the input data items
$x_{j-(i'-i)}$.

\textbf{Farm} The farm skeleton (denoted by the prefix operator
``\farm'') models functional replication. Given a farm skeleton
$\farm(\texttt{\skel})$ the function computed is given by
$\semf[\farm(\texttt{\skel})] = {\cal F}[\texttt{\skel}]$ and the type
of the resulting function will be $\semf[\farm(\texttt{\skel})] :
\alpha \to \beta $ provided that $\semf[\texttt{\skel}] : \alpha \to
\beta$. 
The farm skeleton has to be interpreted as identity, from the strict
viewpoint of its functional semantics.  Its effectiveness derives from
its parallel semantics: in the computation of a farm skeleton,
parallelism is exploited as the computations relative to different
(available) items of the input data stream can all be performed in
parallel.  Therefore given a farm skeleton such as
$\farm(\texttt{\skel})$ and an input stream such as \streaml{x}{1}{n}
the computations $\semf[\skel](x_i)$ and $\semf[\skel](x_j)$ ($i \neq
j$) can be performed in parallel once the data items $x_i$ and $x_j$
are available on the input stream.

Using the skeleton framework described above, skeleton expressions
(i.e. programs) such as the following may be written:
\begin{center}
\framebox{
  \begin{minipage}{0.9\linewidth}
    \texttt{Threshold = seq  }\textit{$\ldots\langle$C code here$\rangle\ldots$}\texttt{  endseq}

    \texttt{Contour = seq  }\textit{$\ldots\langle$C code here$\rangle\ldots$}\texttt{  endseq}

    \texttt{Recognize = seq  }\textit{$\ldots\langle$C code here$\rangle\ldots$}\texttt{  endseq}

    \farm (\texttt{Threshold} \pipe\ \texttt{Contour} \pipe\
    \texttt{Recognize}) 
  \end{minipage}
}
\end{center}
\noindent Assuming that the C code in sequential skeleton \texttt{Threshold}
processes \texttt{Bitmaps} in order to keep black the pixels whose
color value is above a given threshold ($\semf[\texttt{Threshold}] :
\texttt{Bitmap} \to \texttt{Bitmap}$), that the C code in the
\texttt{Contour} skeleton processes black and white bitmaps in order
to recognize contour lines ($\semf[\texttt{Contour}] : \texttt{Bitmap}
\to \texttt{Contour}[]$) and, eventually, that the C code in
\texttt{Recognize} recognizes printable characters out of a set of
contour lines ($\semf[\texttt{Recognize}] : \texttt{Contour}[] \to
\texttt{char}[]$), the skeleton expression \farm (\texttt{Threshold}
\pipe\ \texttt{Contour} \pipe\ \texttt{Recognize}) for each
\texttt{Bitmap} of an input data stream places onto an output data
stream a \texttt{char} vector holding all the characters recognized in
the \texttt{Bitmap}. Parallelism is exploited both by overlapping the
computations of \texttt{Threshold}, \texttt{Contour} and
\texttt{Recognize} relative to one of the \texttt{Bitmaps} appearing
onto the input stream (this is because of the pipeline skeleton)
\textit{and} by performing in parallel the computations relative to
the different \texttt{Bitmap}s appearing onto the input stream (this
is because of the outermost farm skeleton).

\subsection{Stream parallel rewriting rules}
\label{sec:rewriting}
Many equivalences can be established between skeleton expressions
involving the stream parallel skeletons defined in Section
\ref{sec:skeletons}. 
Those equivalences can be proved by looking at the functional semantics
of the skeletons involved. As an example, given the two skeleton
expressions $\skel$ and $\farm(\skel)$, we can immediately see that
both compute the same function $\semf(\skel)$, due to the fact that
farm represents identity from the functional semantics viewpoint.
Therefore we can conclude that, from the point of view of the results
computed by the programs denoted by the two expressions, $\skel \equiv
\farm(\skel)$.
This, in turn, leads to the two rewriting rules \textbf{Fe} (farm
elimination) and \textbf{Fi} (farm introduction) of Figure
\ref{fig:rewritingrules}: the \textbf{Fi} rule allows farms to be
introduced on top of any skeleton expression, whereas \textbf{Fe} rule
allows farms to be removed from the top of any skeleton expression.
Both rules preserve the functional semantics. Instead, parallel
semantics is not preserved, since different kinds and, consequently,
different amounts of parallelism are exploited in the left-hand 
and right-hand side. 

Figure \ref{fig:rewritingrules} presents some of the rewriting rules
we can prove for the skeleton set taken into account in this paper
(the acronyms on the right stand for farm introduction and elimination
(\textbf{Fi}, \textbf{Fe}), pipeline and sequential composition
associativity (\textbf{Pas1}, \textbf{Pas2}, \textbf{SCas1},
\textbf{SCas2}), sequential composition elimination and introduction
(\textbf{Se}, \textbf{Si}), pipe collapse and sequential composition
expansion (\textbf{Coll}, \textbf{Expd})).  These rules state that
farms can be introduced or removed from skeleton expressions, that
pipeline and sequential composition skeletons are associative, that
the sequential composition of a single sequential skeleton is
equivalent to the sequential skeleton itself, and that pipeline and
sequential skeletons can be freely interchanged. 
Due to pipeline and sequential composition associativeness, we will
simplify in the following $(\ldots ((\iota_1;\iota_2);\iota_3); \ldots
\iota_n)$ by $\Seqcomp{\iota}{1}{n}$.

These transformations can be performed without affecting the result
computed by the program, but affecting the amount and the kind of
parallelism exploited and, therefore, the performance eventually
achieved.
Despite their simplicity (we do not prove their validity here, but the
proofs are straightforward), these rules can be effectively used to
rewrite any stream parallel skeleton expression into a ``normal form''
expression (defined in Section \ref{sec:collapsing}) which will be
proven to be more efficient that the non-normal, original one.

\begin{figure}
 \begin{center}
   \framebox{
    \begin{minipage}{0.9\linewidth}
      \vspace*{1ex}
      $\skel \to \farm(\skel)$ \hfill \textbf{(Fi)}
      
      $\farm(\skel)\to \skel $ \hfill \textbf{(Fe)}

      $(\skel_1 \pipe (\skel_2 \pipe \skel_3)) 
      \to 
      ((\skel_1 \pipe \skel_2) \pipe \skel_3)$
      \hfill \textbf{(Pas1)}
      
      $((\skel_1 \pipe \skel_2) \pipe \skel_3)
      \to 
      (\skel_1 \pipe (\skel_2 \pipe \skel_3)) $
      \hfill \textbf{(Pas2)}
      
      $(\stage_1 \seqcomp (\stage_2 \seqcomp \stage_3)) 
      \to
      ((\stage_1 \seqcomp \stage_2) \seqcomp \stage_3)$ 
      \hfill \textbf{(SCas1)}
      
      $((\stage_1 \seqcomp \stage_2) \seqcomp \stage_3)
      \to
      (\stage_1 \seqcomp (\stage_2 \seqcomp \stage_3))$ 
      \hfill \textbf{(SCas2)}

      $\seqcomp(\stage) \to  \stage$ \hfill \textbf{(Se)}
      
      $\stage \to \seqcomp(\stage)$ \hfill \textbf{(Si)}
      
      
      
      $(\stage_1 \pipe \ldots \pipe \stage_k)
      \to
      (\stage_1 \seqcomp \ldots \seqcomp \stage_k)$ 
      \hfill \textbf{(Coll)}
      
      $(\stage_1 \seqcomp \ldots \seqcomp \stage_k)
      \to
      (\stage_1 \pipe \ldots \pipe \stage_k)$ 
      \hfill \textbf{(Expd)}
      
      \vspace{1ex}
    \end{minipage}
    }
  \end{center}\vspace*{-1em}
  \caption{\textit{Rewriting rules}}
  \vspace{-2ex}
  \label{fig:rewritingrules}
\end{figure}

\subsection{Performance estimation}
\label{sec:performance}
We assume that programs written with the skeletons of Section
\ref{sec:skeletons} are compiled to parallel object code using
template based compiling tools, such as those used in both \pppl\ and
\skie\ 
\cite{orlando-grosso,skie-userman-98,libro-susanna,skie-marcov}. Such
kind of tools are based on the existence of a predefined, parametric
and efficient process network for each one of the skeletons. These
parametric process networks, the implementation templates, are used to
implement the skeleton instances. Here, in particular, we assume that
each implementation template is a parametric (in the number of
processing nodes used) process network with a single input and a
single output ``point''. That is, we assume that there is a unique
place where data of the input stream are taken and another unique
place where data of the output stream are placed by the template
process network (these ``places'' can be channels, memory locations,
etc.).
Although this assumption may look like to be restrictive, it perfectly
models the concept of data stream, which is a \textit{single}
entity hosting data items available at different,
consecutive times.  Furthermore, this assumption allows effective
implementation template composition to be performed, which in turn is
necessary to implement full skeleton nesting \cite{libro-susanna}.
This is also the assumption currently made in the compiling tools of
both \pppl\ and \skie, the skeleton based parallel programming
environments currently being developed in Pisa.


In order to estimate the performance of our skeleton programs, we need
some performance model associated to the implementation templates we
assume to use to implement the skeletons on the target architecture.
Such performance models should reflect the performance achieved when
implementing a particular process network on a particular target
architecture, however, and therefore they are strongly related to the
templates used and to the target machine used.
Instead, we are interested in the general properties of the skeletons,
and therefore here we try to abstract the performance models of the
templates from templates peculiarities and from target architecture
features.  Eventually, we come up with ``ideal'' performance models
that have to be intended as asymptotic models, i.e. models that
represent the \textit{best} performance that can be achieved by a
template implementing the skeleton under the assumptions made in this
work.  This because we are interested in the evaluation of the lower
bounds, rather than in the precise modeling of the template
performance.

In this work we consider the service time of a template as the
performance measure to be optimized. The service time is the time
occurring between the delivery of two distinct, consecutive result
data items onto the output stream. It can also be defined as
the time necessary to the first process of the template to accept and
process a new input data item \footnote{This measure is different from
  the completion time, i.e. the time occurring between the arrival of
  the first data item of the input stream to the template and the
  delivering of the last result data item onto the output stream.}.
%


We introduce now the ideal performance models for the implementation
templates of our skeletons.  We denote by $\service{\skel}$ the
service time of the template implementing the skeleton $\skel$, and by
$\tseq{\stage}$ the average amount of time spent in computing the code
embedded by the \seq\ skeleton $\stage = \seq(\ldots)$. We denote by
$\tin{\skel}$ and $\tout{\skel}$ the time spent by the template
implementing \skel\ in receiving an input data item and delivering an
output data item from/to a distinct processing node. These parameters
can be used to model either the overhead associated with a
communication or the actual (total) time spent in the communication,
depending on the parallelism degree of the target architecture
processing element (i.e.  on the presence of some kind of independent
communication processor).

\noindent \textbf{Sequential skeleton:} $\service{\stage} 
= \seqservice{\stage}$ i.e. the service time of a sequential skeleton
should be at least equal to the time spent in receiving the input data
plus the time spent in executing the code, plus the time spent in
delivering the result data.

\noindent \textbf{Sequential composition skeleton:}
$\service{\Seqcomp{\stage}{1}{k}} = $ $\tin{\stage_1} +
\tout{\stage_k} + \sum_{i=1}^{k} \tseq{\stage_i}$ i.e. the service
time of a sequential composition of sequential skeletons should be at
least equal to the time spent in gathering parameters and delivering
results plus the time spent in the computation of each one of the
\seq\ skeletons involved.

\noindent \textbf{Pipeline:} $\service{\Pipe{\skel}{1}{k}} =
\pipeservice{\skel}{1}{k}$\footnote{We denote by $\max\{ \ldots \}$
  and $\min\{ \ldots \}$ the maximum and the minimum of a set,
  respectively.} i.e. the service time of a pipeline should be at least
equal to the maximum of the service times of its stages \cite{ni1}.

\noindent \textbf{Farm:} $\service{\farm{\skel}} = 
\farmservice{\skel}$, denoting the fact that a farm template with
single input/output points cannot serve requests in a time shorter
than the time spent in receiving an input or delivering an output data
item.  Furthermore, it does not make sense to use a farm when the time
spent in delivering a data item is greater that the time spent in
computing the result of that data item, as in this case the results
could be computed with better service time without using the farm
\cite{natug}.  It's worthwhile pointing out that this kind of
performance modeling of the farm corresponds to the assumption that a
farm template is basically a three stage pipeline: the first stage
gathers data items from the input data stream, the second (parallel!)
stage computes the farm results and the third stage gathers the
results from the second one and delivers them on the output stream.

\section{Collapsing skeletons\vspace*{-1ex}}
\label{sec:collapsing}
We first define a ``normal form'' of stream parallel skeleton
compositions, then we will show how normal forms of skeleton
compositions always achieve an equal or better performance (in terms
of the service time) w.r.t. the equivalent, non-normal form skeleton
compositions, despite their simpler structure in terms of the skeleton
nesting used. 

Given any skeleton composition $\Delta$, we define \frangia{\Delta} to
be the ordered list of all the sequential portions of code in
$\Delta$.  Formally, we define \frangia{\Delta} by induction on the
structure of $\Delta$ as follows:

\begin{itemize}
\item $\frangia{\stage} = \stage$
\item \vspace{-0.8em} $\frangia{\Seqcomp{\stage}{1}{k}} = [\llista{\stage}{1}{k}]$
\item \vspace{-0.8em} $\frangia{\farm(\skel)} = \frangia{\skel}$
\item \vspace{-0.8em} $\frangia{\skel_1 \pipe \skel_2} =
  \texttt{append}(\frangia{\skel_1},\frangia{\skel_2})$
\end{itemize}

\noindent We define the \textbf{normal form} of a generic skeleton composition
$\Delta$, denoted by $\normale{\Delta}$, as\footnote{We use the
  sequential composition operator in prefix form, in this case,
  assuming that $\seqcomp(\stage_1,\ldots,\stage_k) \equiv \stage_1
  \seqcomp \ldots \seqcomp \stage_k$.}
  
\[ \normale{\Delta} =
\farm(\seqcomp(\frangia{\Delta})) \]


\noindent It is easy to prove the following statement:
\begin{teo}
  Given any skeleton composition $\Delta$, $\semf[\Delta] = \semf[\normale{\Delta}]$
\end{teo}

\paragraph{Proof} The proof is by induction 
on the structure of the skeleton program.

\noindent \textbf{Base case} The program just contains a sequential
skeleton or a sequential composition one. In this case  
either $\Delta
= \stage = \texttt{seq(...)}$ and therefore 
$\stage \rufreccia{Si} \seqcomp(\stage) \rufreccia{Fi} 
\farm(\seqcomp(\stage))$ 
or $\Delta =
\Seqcomp{\stage}{1}{k} $ and therefore $\Seqcomp{\stage}{1}{k}
\ \rufreccia{Fi}\  \farm(\Seqcomp{\stage}{1}{k})$

\noindent \textbf{Inductive cases} The program is either a pipe or a
farm of normal form skeletons. In this case 
\begin{itemize}
\item 
either $\Delta =
(\farm(\Seqcomp{\iota}{11}{1n_1})) \pipe \ldots \pipe
(\farm(\Seqcomp{\iota}{k1}{kn_k}))$, \\
therefore\\
$  (\farm(\Seqcomp{\iota}{11}{1n_1})) \pipe \ldots \pipe
   (\farm(\Seqcomp{\iota}{k1}{kn_k}))
   \ \rufreccia{Fe}$ \hfill\\
$
   (\Seqcomp{\iota}{11}{1n_1}) \pipe$ $\ldots$ $\pipe
   (\Seqcomp{\iota}{k1}{kn_k})
   \ \rufreccia{Coll}$ \hfill\\
$
   (\Seqcomp{\iota}{11}{1n_1})
   \seqcomp \ldots \seqcomp (\Seqcomp{\iota}{k1}{kn_k})
   \ \rufreccia{SCas}$ \hfill\\
$
   (\Seqcomp{\iota}{11}{1n_1};\ldots;\Seqcomp{\iota}{k1}{kn_k})
   \ \rufreccia{Fi}$ \hfill\\
$
   \farm(\Seqcomp{\iota}{11}{1n_1}
   \seqcomp$ $\ldots$ $\seqcomp \Seqcomp{\iota}{k1}{kn_k})$, \hfill
\item 
or $\Delta =
  \farm(\farm(\Seqcomp{\stage}{1}{n}))$ and therefore \hfill\\
  $\farm(\farm(\Seqcomp{\stage}{1}{n}))\ \rufreccia{Fe}\ 
  \farm(\Seqcomp{\stage}{1}{n})$ \hfill
\end{itemize}

\subsection{Collapsing effects}
\label{sec:effects}
%
Using the abstract performance models for templates introduced in
Section \ref{sec:performance}, we can prove the following result,
which is the main result discussed in this paper.

\begin{teo}
  Given any stream parallel composition $\Delta$, such that $\forall
  \stage \in \frangia{\Delta}$ $\tin{\stage} < \tseq{\stage}$ and
  $\tout{\stage} < \tseq{\stage}$, then
  $\service{\normale{\Delta}} \leq \service{\Delta}$
\end{teo}

\noindent \textbf{Proof}
We prove the statement by induction on the structure of $\Delta$.
  
\noindent \textbf{Base cases} Either $\Delta =
\stage$ or $\Delta = \Seqcomp{\stage}{1}{k}$ and therefore:
\begin{itemize}
\item 
$\Delta = \stage$\\
  $\service{\Delta} = \min\{\service{\Delta}\} \geq $
  $\min \{ \max \{ \tin{\stage}, \tout {\stage} \} , \service{\stage} \} = $
  \hfill\\
  $\service{\farm(\seqcomp(\stage))} = $ 
  $\service{\normale{\Delta}}$
\item 
$\Delta = \Seqcomp{\stage}{1}{k}$\\
  $ \service{\Delta} = $
  $ \min \{ \service{\Delta} \} \geq $\hfill\\
  $ \min \{ \max \{ \tin{\stage_1} , \tout{\stage_k} \} , \service{\Delta} \} = $\hfill\\
  $ \min \{ \max \{ \tin{\stage_1} , \tout{\stage_k} \} , \service{\llista{\stage}{1}{k}} \} = $\hfill\\
  $ \service{\farm(\llista{\stage}{1}{k})} = $ 
  $ \service {\normale {\Delta}}$ 
\end{itemize}

\noindent \textbf{Inductive cases} Either $\Delta = \farm \skel$ or
$\Delta = \skel_1 \pipe \skel_2$ and therefore:

\begin{itemize}
\item $\Delta = \farm \skel$. In this case $\service{\skel} \geq
  \service{\normale{\skel}}$ by induction hypothesis. Furthermore, we
  assume that $\frangia{\skel} =  \lista{\iota}{k}$ and therefore:

  $ \service{\Delta} = \service{\farm{\skel}} = $
  $ \farmservice{\skel} \geq $ \hfill\\
  $ \min \{ \max \{ \tin{\skel}, \tout{\skel}\}, 
  \service{\normale{\skel}} \} = $\hfill\\
  $ \min \{ \max\{\tin{\iota_1}, \tout{\iota_k}\}, 
  \service{\normale{\skel}} \} = $\hfill\\
  $ \min \{  \max\{\tin{\iota_1}, \tout{\iota_k}\}, 
  \min \{ \max \{\tin{\stage_1},$\hfill\\
  \hspace*{6ex} $\tout{\stage_k}\}, \service{\Seqcomp{\stage}{1}{k}} \}\} = $ \hfill\\
  $\min \{ \max \{ \tin{\iota_1}, \tout{\iota_k}\},$ 
  $\max \{\tin{\stage_1},
  \tout{\stage_k}\},$ \hfill\\ 
  \hspace*{6ex} $\service{\Seqcomp{\stage}{1}{k}}\} = $ \hfill\\
  $\min \{ \max \{\tin{\iota_1}, \tout{\iota_k}\}, 
  \service{\Seqcomp{\stage}{1}{k}}\} = $ \hfill\\
  $ \service{\farm{(\Seqcomp{\stage}{1}{k})}} = 
  \service{\normale{\farm{\skel}}} $

\item $\Delta = \skel_1 \pipe \skel_2$ (with $\frangia{\skel_1} =
  \iota_{11}, \ldots, \iota_{1n_1}$ and $\frangia{\skel_2} =
  \iota_{21}, \ldots, \iota_{2n_2}$). In this case, $\service{\skel_i}
  \geq \service{\normale{\skel_i}}, i \in \{ 1,2 \}$ by induction
  hypothesis:

  $\service{\skel_1 \pipe \skel_2} = $
  $\max \{ \service{\skel_1}, \service{\skel_2} \} \geq $ \hfill\\
  $\max \{ \service{\normale{\skel_1}}, \service{\normale{\skel_2}} \}
  = $ \hfill\\
  $\max \{ \service{\farm(\iota_{11};\ldots; \iota_{1n_1})}, 
           \service{\farm(\iota_{21};\ldots; \iota_{2n_2})} \} $ = \hfill\\
  $\max \{ 
  \min \{ \max \{\tin{\stage_{11}} ,\tout{\stage_{1n_1}}  \} ,
  \sum_{i=11}^{1n_1}\tseq{\stage_i}  \}, $ \hfill\\
  \hspace*{6ex} $\min \{ \max \{ \tin{\stage_{21}}, \tout{\stage_{2n_2}}\} ,
  \sum_{i=21}^{2n_2}\tseq{\stage_i}  \} \} = $ \hfill\\
  $\max \{ \max \{\tin{\stage_{11}} ,\tout{\stage_{1n_1}}  \} , $
  $\max \{ \tin{\stage_{21}},   \tout{\stage_{2n_2}}\}\}  = $\hfill\\
  $\max \{\tin{\stage_{11}} ,\tout{\stage_{1n_1}} , 
  \tout{\stage_{2n_2}}\} \geq $\hfill\\
    $\max \{\tin{\stage_{11}},  \tout{\stage_{2n_2}}\} \geq $\hfill\\
  $  \min \{ \max \{ \tin{\stage_{11}}, \tout{\stage_{2n_2}} \}\} \geq $
  \hfill\\
  $\min \{ \max \{ \tin{\stage_{11}}, \tout{\stage_{2n_2}} \}, 
  \sum_{i=11}^{1n_1}\tseq{\stage_i}+$\hfill\\
  \hspace*{6ex} $\sum_{i=21}^{2n_2}\tseq{\stage_i}\} = $\hfill\\
  $\service{\farm(\Seqcomp{\stage}{11}{2n_2})} = $
  $\service{\normale{\skel_1 \pipe \skel_2}} $\hfill
\end{itemize}

This result tells us that any time we have a stream parallel
composition in a larger skeleton composition
we can rewrite it in normal form and expect that performance (service
time) is either the same or better than the original one. The
assumptions on the input and output times ($\forall \stage \in
\frangia{\Delta}$ $\tin{\stage} < \tseq{\stage}$ and $\tout{\stage} <
\tseq{\stage}$) are not restrictive in that every time we have a
module whose latency is smaller than the time spent to feed the module
with new data to be computed and to extract results out of the module,
the parallel module can be conveniently eliminated and a sequential
module can be used instead.

\begin{figure*}
\begin{center}
{
\begin{tabular}{|l|ccccccc|} \hline
\textbf{TableA}  & $\stage_1 ; \stage_2$ & $\farm(\stage_1 ;
\stage_2)$ & $\farm(\farm(\stage_1) \pipe \farm(\stage_2))$ &
$(\farm(\stage_1) \pipe \farm(\stage_2))$ & $\farm(\stage_1 \pipe 
\stage_2)$ & $ \farm(\stage_1) \pipe \stage_2$ & 
$\stage_1 \pipe \farm(\stage_2)$ \\ \hline
$T_s$  &6.03     &0.33     &0.35        &0.37
  &0.35    &1.08      &4.98 \\
$T_c$      &1207.76   &71.11      &76.60      &81.00
  &74.64   &222.04    &1003.75\\ 
\#PE &1      &24     &44     &24     &34     &9      &7\\
$\epsilon\ (\%)$ &     &75.60   &38.85   &66.99
  &50.71  &62.05   &17.29\\  \hline
\end{tabular}}\\[1em]    
{
\begin{tabular}{|l|ccccccc|} \hline
\textbf{TableB}  & $\stage_1 ; \stage_2$ & $\farm(\stage_1 ; 
\stage_2)$ & $\farm(\farm(\stage_1) \pipe \farm(\stage_2))$ &
$(\farm(\stage_1) \pipe \farm(\stage_2))$ & $\farm(\stage_1 \pipe 
\stage_2)$ & $ \farm(\stage_1) \pipe \stage_2$ & 
  $\stage_1 \pipe \farm(\stage_2)$ \\ \hline
$T_s$  & 6.03    &   0.39  &      1.30  &      0.72& 0.43  &      1.12 &      4.99\\
 $T_c$     & 1207.76     &84.50    &    286.62  &    151.67  &    91.43    &   230.35    &  1004.69\\
 \#PE &   1   &    20     & 20   &   20  &    20 &     20  &    20\\

$\epsilon\ (\%)$ & &              75.52  &   23.08  & 41.93 &   69.56   &
26.88 &   6.04\\ \hline
\end{tabular}}    

\caption{\textit{Normal form vs. non-normal forms. \textbf{Table A}:
optimal number of 
      processing elements for each run \textbf{Table B}: same number of
      processing elements for each run 
        ($T_s$: service time, $T_c$:
      completion time, $\#PE$: number of processing elements used,
      $\epsilon$: efficiency)}}
    \label{fig:experimental3}
  \end{center}
\end{figure*}

Concerning the effects of collapsing stream parallel skeleton compositions to
normal form on the resources used, a different kind of reasoning has to
be performed.
The farm introduced by the normal form compute ``heavy'' functions,
the functions obtained by merging all the sequential stages of the
original composition. Therefore, in order to be effective, a large
number of parallel processes have to be included in the farm template,
in such a way that, every time a new data item is available to schedule
a new parallel computation, a process happens to be idle, ready to
start this new computation. This, in turn, allows to achieve the
asymptotic performance modeled by performance formulas of Section
\ref{sec:performance}. The number of parallel processes used in the
normal form farm should therefore be something like
$\frac{\service{\llistapv{\iota}{1}{k}}}
{\max\{\tin{\iota_1},\tout{\iota_k}\}}$.
Whether or not this amount of processing resources (parallel processes)
is greater than the amount of resources needed to implement the
corresponding non-normal form is not known, at the moment, but we are
currently looking for a result, similar to the one discussed above for
the service time, that could assess the relationships between the
resources used in the normal and non-normal forms of stream parallel
skeletons compositions.
We expect that normal form programs can be implemented with a smaller
amount of resources dedicated to the skeleton run time support,
because of the simpler skeleton structure of normal form programs with
respect to non-normal form ones. This should lead to smaller overheads
in the execution of normal forms and, in general, to a better
utilization of the processing elements at hand.\footnote{We expect to
  achieve better service times \textit{and}, at the same time, better
  efficiency.}
Preliminary experimental results (shown in Section
\ref{sec:experiments}) seem to validate this kinds of reasoning.

\subsection{Experimental results}
\label{sec:experiments}
We performed some experiments aimed at validating the theoretical
results discussed in Section \ref{sec:effects}. We used two skeleton
compilers, Anacleto \cite{anacleto-australia}, the Pisa prototype
\pppl\ compiler, and the
skeleton compiler integrated within \skieenv\ \cite{skie-parcom}. Both
compilers handle a skeleton language including the stream parallel
skeletons discussed in Section \ref{sec:skeletons}, and both use
implementation templates such as those discussed in Section
\ref{sec:performance}.

In order to check whether normal form programs perform better than
equivalent, non-normal form ones, we performed multiple runs of normal
and non-normal versions of the same programs on a cluster of 10
Pentium II PCs interconnected by a 100Mbit switched Ethernet and on a
Fujitsu AP1000 parallel machine.

\begin{figure*}
  \begin{center}
    \leavevmode
    \hspace*{-0.2cm}\includegraphics[scale=0.7]{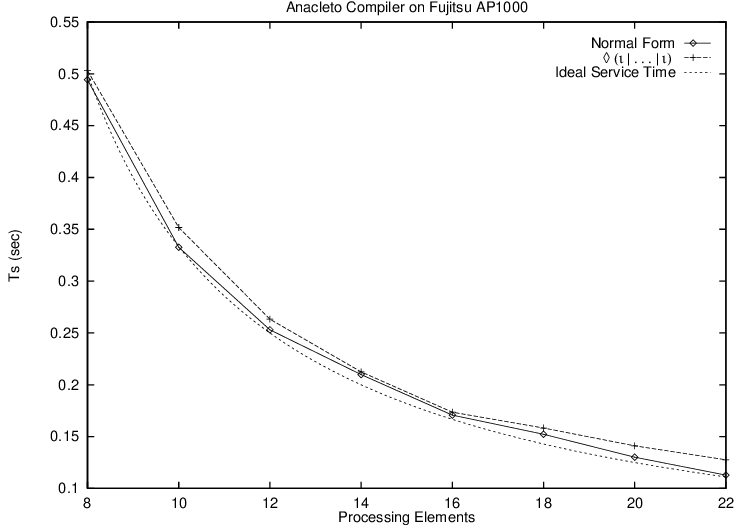}
    \includegraphics[scale=0.7]{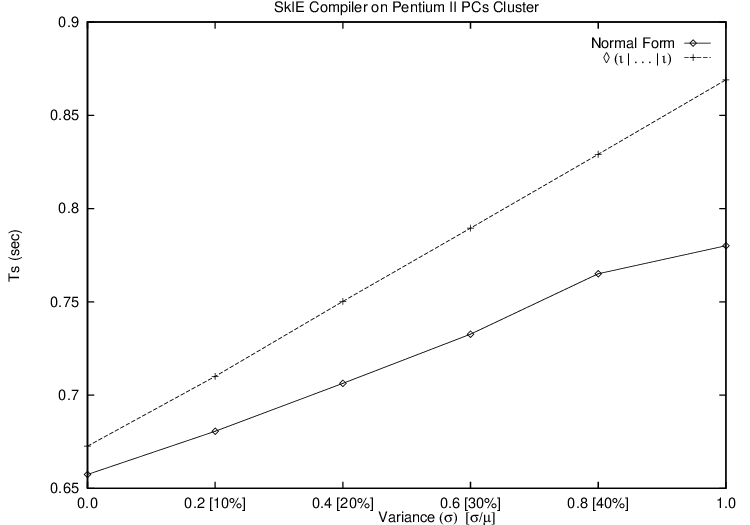}
   \caption{\textit{Experimental results: service time (in seconds)
    vs. number of 
    processing elements used (left) service times (in seconds) vs. variance of
    sequential skeleton time (right)}}
    \label{fig:experimental1}
  \end{center}
\end{figure*}

First of all, we looked at the behavior of the different, equivalent
forms of a program, with respect to performance.  Table A of Figure
\ref{fig:experimental3} summarizes the service and completion time,
the resource usage count and the efficiency measured when running
different forms (equivalent skeleton compositions) of the same
program.  In this case, the program is a two stage pipeline with the
first stage taking five times the time taken by the second one to
compute a result.\footnote{The latencies of sequential computations
are randomly chosen in accordance with a normal distribution
$N(\mu,\sigma)$ with variance $\sigma = 0.6$.}
All the different versions compute a stream of 200 input
tasks (on a Fujitsu AP1000).
In this first set of runs, we used the exact amount of resources
(processing elements) that maximizes the run performance. This amount
of resources can be derived by the performance models associated
to the implementation templates used to implement the skeletons and
depends, of course, on the kind of skeleton nesting used within the
program.
The normal form (second column, in the table) run takes less time to
complete ($T_C$), delivers a better service time ($T_S$) and presents
a better efficiency (computed on the service time) than the others,
semantically equivalent forms of the program. In particular, the
normal form uses the same amount of processing elements of the pipe of
farms version, but this version of the program is slower.

In Table B of Figure \ref{fig:experimental3}, we summarize the times taken when
the different versions of the programs are run using the same amount
of processing elements. 
The amount of processing elements used has been chosen to be slightly
smaller of the ``optimal'' amount of processing elements required by
the normal form and farm of pipeline programs.
In this case, the advantage coming from the
usage of the normal form is more evident. This is due to the
``better'' usage of the available processing elements by normal
form (i.e. to the smaller overheads introduced in computations and to
the smaller number of processing elements dedicated to the execution
of the bare skeleton/template run time support).

We also considered programs that are expected to pay a minor
performance slowdown with respect to the normal form programs: we
considered skeleton compositions in the form
$\Farm{\Pipe{\iota}{1}{k}}$ vs. the equivalent normal forms
$\Farm{\Seqcomp{\iota}{1}{k}}$. In both cases the mean load of all
sequential computations is equal and fixed. We expected minimum
differences between the service times of first and second form of
these programs, because of the poor structuring of the first form.
Furthermore, we expected that normal form is not sensibly better than
the other form when the difference between times spent in sequential
computations is low, as in this case the load balancing effects of
normal form farm are poorer.
Figure \ref{fig:experimental1} (left) plots the service time of a
$\Farm{\Pipe{\iota}{1}{k}}$ program in function of the parallelism
degree (the number of processing elements used, actually) vs. the
service time of the equivalent normal form program (times taken on a
Fujitsu AP 1000).  The ideal service time (i.e. the service time
computed by using the performance models of Section
\ref{sec:performance}) of normal form program is also plotted. The
normal form times are better than non-normal form times even if the
load of all the sequential computations is exactly the same (and this
should be the best condition for farms of pipelines with respect to
the corresponding normal form: the farm of sequential composition
skeleton) and, furthermore, normal form times are very close to the
ideal ones.

In order to evaluate the effect of the local load imbalances on normal
and non-normal form programs, we run programs with a variable variance
in the sequential latencies.  Figure \ref{fig:experimental1} (rigth)
plots the service times of programs when the latencies of sequential
computations are randomly chosen in accordance with a normal
distribution $N(\mu,\sigma)$ (times taken on the Linux Pentium II/Fast
Ethernet cluster). Again, the normal form times are better than those
achieved with the equivalent non-normal form programs.  Moreover, due
to the better load balancing capabilities of the normal form, the gap
between the performances grows as the unbalancing of sequential
latencies grows.
We obviously expect that even better results can be measured in case
of more structured programs, as in this case normal form leads to a
sensible decrease in the run time support code that has to be executed
to distribute data amongst the processing elements involved in the
parallel execution of skeleton code.

\section{Related work and conclusions\vspace*{-1ex}}  
\label{sec:conclusions}
Rewriting techniques with some kind of associated cost calculus have
been developed by different research groups working on skeletons,
parallel functional programming and general structured parallel
programming.

Papers have been published demonstrating the potential usage of
a particular cost calculus in the evaluation of program
transformations \cite{bsp-optim,jay98}. 
Other researchers developed transformation techniques aimed at
improving specific data parallel computations \cite{gorlatch97},
mostly derived from the Bird-Merteens parallel functional programming
framework \cite{bird-grosso}.
However, most of these works just deal with data parallel computations
and the associated transformations.

Both the Darlington group at the Imperial College and the authors'
group working on \pppl\ discussed transformation rules involving both
stream (control) parallel and data parallel skeletons
\cite{darlington-parle,orlando-grosso,modelli-aldinuc-coppola}.  Even
in these works, however, no idea of ``best'' or ``normal form''
skeleton composition such as the one discussed here has been
presented.

The main result discussed in this paper is that any arbitrary
composition of stream parallel skeletons can be rewritten into an
equivalent ``normal form'' skeleton composition. Such normal form
skeleton composition computes the same results computed by the
original program, delivering a service time which is better or equal to
the service time of the original program. This result has been derived
taking into account ideal performance models, i.e. not taking into
account all the overheads associated with the exploitation of nested
skeleton programs, which are quite hard to measure within the abstract
cost model we devised for the skeleton implementations.  Therefore it
has to be considered ``optimistic'' in the sense that normal forms are
subject to a smaller amount of overhead with respect to highly nested,
equivalent, non-normal forms.

Experimental results demonstrated that normal form programs deliver a
better service time than the equivalent, non-normal form ones. All the
experiments we performed on different machines, with different
compiler and different programs showed that the normal form programs
run faster than the equivalent, non-normal ones even in those cases
where we expected the behavior to be quite close (e.g.  farm of
pipeline of balanced sequential stages).
Furthermore, in those cases where the load imbalances in the
sequential computations affect the load balancing features of
non-normal form program implementations, normal forms achieve a
sensibly better service time.

Currently, we are investigating two different ways to extend the
results discussed in this work.  On the one hand, we are trying to
evaluate the relationship between the number of resources (processing
elements) needed to implement non-normal forms and the equivalent
normal form programs.
Experimental results show that the amount of resources needed to run
normal form programs is close (usually smaller) to the amount of
resources needed to run the equivalent non normal form programs
achieving the best performance.
On the other hand, we are currently investigating whether or not some
kind of normal form can be found even in case that data \textit{and} stream
parallel skeletons are taken into account.

\noindent \textbf{Acknowledgments} We wish to thank Prof. Christian Lengauer
for his useful comments to the first version of our work, and the 
Dept. of Computing, Imperial College, London, for the machines access.
\vspace{-2ex}


{\footnotesize
\bibliographystyle{unsrt}
\bibliography{mdbib}

\begin{thebibliography}{10}

\bibitem{cole-th}
M.~Cole.
\newblock {\em {Algorithmic Skeletons: Structured Management of Parallel
  Computations}}.
\newblock Research Monographs in Parallel and Distributed Computing. Pitman,
  1989.

\bibitem{darlington-parle}
J.~Darlington, A.~J. Field, P.G. Harrison, P.~H.~J. Kelly, D.~W.~N. Sharp,
  Q.~Wu, and R.~L. While.
\newblock {Parallel Programming Using Skeleton Functions}.
\newblock In M.~Reeve A.~Bode and G.~Wolf, editors, {\em PARLE'93 Parallel
  Architectures and Langauges Europe}. Springer Verlag, June 1993.
\newblock LNCS No. 694.

\bibitem{fgcs-firenze}
M.~Danelutto, R.~Di Meglio, S.~Orlando, S.~Pelagatti, and M.~Vanneschi.
\newblock A methodology for the development and support of massively parallel
  programs.
\newblock {\em Future Generation Computer Systems}, 8(1--3):205--220, July
  1992.

\bibitem{bacs}
H.~Burkhart and S.~Gutzwiller.
\newblock {Steps Towards Reusability and Portability in Parallel Programming}.
\newblock In K.~M. Decker and R.~M. Rehmann, editors, {\em Programming
  Environments for Massively Parallel Distributed Systems}, pages 147--157.
  Birkhauser, April 1994.

\bibitem{Darlington1996}
P.~Au, J.~Darlington, M.~Ghanem, Y.~Guo, H.W. To, and J.~Yang.
\newblock Co-ordinating heterogeneous parallel computation.
\newblock In L.~Bouge, P.~Fraigniaud, A.~Mignotte, and Y.~Robert, editors, {\em
  Europar '96}, pages 601--614. Springer-Verlag, 1996.

\bibitem{skie-parcom}
B.~Bacci, M.~Danelutto, S.~Pelagatti, and M.~Vanneschi.
\newblock {SkIE: a heterogeneous environment for HPC applications}.
\newblock Parallel Computing, to appear, 1999.

\bibitem{skie-marcov}
M.~Vanneschi.
\newblock {PQE2000: HPC tools for industrial applications}.
\newblock {\em IEEE Concurrency}, (4):68--73, 1998.

\bibitem{orlando-grosso}
B.~Bacci, M.~Danelutto, S.~Orlando, S.~Pelagatti, and M.~Vanneschi.
\newblock {P$^3$L: A Structured High level programming language and its
  structured support}.
\newblock {\em Concurrency Practice and Experience}, 7(3):225--255, May 1995.

\bibitem{libro-susanna}
S.~Pelagatti.
\newblock {\em {Structured Development of Parallel Programs}}.
\newblock Taylor \& Francis, 1998.

\bibitem{anacleto-australia}
S.~Ciarpaglini, M.~Danelutto, L.~Folchi, C.~Manconi, and S.~Pelagatti.
\newblock {ANACLETO: a template-based P3L compiler}.
\newblock In {\em Proceedings of the PCW'97}, 1997.
\newblock Camberra, Australia.

\bibitem{bird-grosso}
R.~S. Bird.
\newblock Lectures on constructive functional programming.
\newblock In Manfred Broy, editor, {\em Constructive Methods in Computing
  Science}. {NATO ASI} Series, 1988.
\newblock International Summer School directed by F. L. Bauer, M. Broy, E. W.
  Dijkstra and C. A. R. Hoare.

\bibitem{gorlatch97}
S.~Gorlatch and C.~Lengauer.
\newblock {(De)Composition Rules for Parallel Scan and Reduction}.
\newblock In {\em 3rd Int. Conf. on Massively Parallel Programming Models},
  November 1997.
\newblock London.

\bibitem{modelli-aldinuc-coppola}
M.~Aldinucci, M.~Coppola, and M.~Danelutto.
\newblock {Rewriting skeleton programs: how to evaluate the data-parallel
  stream-paralle treadoff}.
\newblock In {\em Proceedings of the International Workshop on Constructive
  Methods for Parallel Programming}, 1998.
\newblock Technical Report, University of Passau, No. MIP-9805.

\bibitem{kessler-svizzera}
C.~W. Kessler.
\newblock {Symbolic Array Data Flow Analysis and Pattern Recognition in
  Numerical Codes}.
\newblock In K.~M. Decker and R.~M. Rehmann, editors, {\em Programming
  Environments for Massively Parallel Distributed Systems}, pages 57--68.
  Birkhauser, April 1994.

\bibitem{skie-userman-98}
B.~Bacci, B.~Cantalupo, P.~Pesciullesi, R.~Ravazzolo, A.~Riaudo, and
  L.~Vanneschi.
\newblock {SKIE: user's guide (version 2.0)}.
\newblock Technical report, QSW Ltd. Rome, Italy, December 1998.

\bibitem{ni1}
C.~T. King, W.~H. Chou, and L.~M. Ni.
\newblock Pipelined data-parallel algorithms: Part {I} -- {C}oncept and
  {M}odeling.
\newblock {\em IEEE Transactions on Parallel and Distributed Systems}, 1(4),
  October 1990.

\bibitem{natug}
B.~Bacci, M.~Danelutto, S.~Pelagatti, S.~Orlando, and M.~Vanneschi.
\newblock {Unbalanced Computations onto a Transputer Grid}.
\newblock In {\em Proceedings of The 1994 Transputer Research and Application
  Conference}, pages 268--282. IOS Press, October 1994.
\newblock Athens, Georgia, USA.

\bibitem{bsp-optim}
D.~B. Skillicorn, M.~Danelutto, S.~Pelagatti, and A.~Zavanella.
\newblock {Optimising Data-Parallel Programs Using the BSP Cost Model}.
\newblock In D.~Pritchard and J.~Reeve, editors, {\em Euro-Par'98 Parallel
  Processing}, pages 698--703. Springer Verlag, 1998.
\newblock LNCS No. 1470.

\bibitem{jay98}
C.~B. Jay, M.~I. Cole, M.~Sekanina, and P.~Steckler.
\newblock {A monadic calculus for parallel costing of a functional language of
  arrays}.
\newblock In C.~Lengauer, M.~Griebl, and S.~Gorlatch, editors, {\em EuroPar'97
  Parallel Processing}, pages 650--661. Springer Verlag, 1997.
\newblock LNCS. No. 1300.

\end{thebibliography}
}


\end{document}